\documentclass[twocolumn]{elsarticle}

\usepackage[dvipdfmx]{color}
\usepackage{hyperref}
\usepackage{graphicx, color}
\usepackage{cleveref}

\hypersetup{ 
    colorlinks=true,
}    
\hypersetup{
    anchorcolor=black, 
    citecolor=cyan,        
    filecolor=cyan,       
    linkcolor=cyan,     
    urlcolor=cyan,
    linkbordercolor={0 0 0},
}
\usepackage{url} 
\usepackage{siunitx}

\usepackage{amsmath, amssymb}
\journal{Journal of \LaTeX\ Templates}

\begin{document}

\twocolumn[{\begin{frontmatter}

\title{A compensated multi-gap RPC with \SI{2}{m}-long strips for the LEPS2 experiment}

\author[kyoto]{K. Watanabe\fnref{myfootnote}}
\author[rcnp]{S. Tanaka}
\author[taiwan]{W. C. Chang}
\author[kirchhoff]{H. Chen}
\author[taiwan]{M. L. Chu}
\author[spain]{J. J. Cuenca-Garc\'ia}
\author[tohoku]{\\ T. Gogami} 
\author[spain]{D. Gonz\'alez-D\'iaz}
\author[kyoto-sangyo]{M. Niiyama}
\author[jasri]{Y. Ohashi}
\author[elph]{H. Ohnishi}
\author[rcnp]{N. Tomida}
\author[rcnp]{M. Yosoi}

\fntext[myfootnote]{Corresponding author.
E-mail address: kenwata@nh.scphys.kyoto-u.ac.jp (K. Watanabe)}



\address[kyoto]{Department of Physics, Kyoto University, Kyoto 606-8502, Japan}
\address[rcnp]{Research Center for Nuclear Physics (RCNP), Osaka University, Ibaraki, Osaka 567-0047, Japan}
\address[taiwan]{Institute of Physics, Academia Sinica, Nankang, Taipei 11529, Taiwan}
\address[kirchhoff]{Kirchhoff Institut f\"ur Physik, Universit\"at Heidelberg, Im Neuenheimer Feld 227, 69120, Germany}
\address[tohoku]{Graduate School of Science, Tohoku University, Sendai, Miyagi 980-8578, Japan}
\address[spain]{Instituto Galego de Fisica de Altas Enerxias (IGFAE), Universidade de Santiago de Compostela, 15782, Spain}
\address[kyoto-sangyo]{Department of Physics, Kyoto Sangyo University, Kyoto 603-8555, Japan}
\address[jasri]{Japan Synchrotron Radiation Research Institute (JASRI), SPring-8, Sayo, Hyogo 679-5198, Japan}
\address[elph]{Research Center for Electron Photon Science (ELPH), Tohoku University, Sendai, Miyagi 982-0826, Japan}

\begin{abstract}
   We have developed multi-gap resistive plate chambers (MRPCs) 
   with $2.5 \times 200$ \si{cm^2} readout strips for the time-of-flight (TOF) detector system 
   of the LEPS2 experiment at SPring-8.
   These chambers consist of  2 stacks and 5 gas gaps per stack, in a mirrored configuration.
   A time resolution of $\sigma \simeq 80$ \si{ps} was achieved for any position within a strip
   (at above 99\% detection efficiency); 
   after performing the time-charge slewing correction, 
   this value could be reduced to \SI{60}{ps}.
   A link between the small contribution of the slewing correction to timing 
   and the suppression of modal dispersion in the detector could be established.
   
\end{abstract}


\end{frontmatter}
}]


\section{Introduction}
   Resistive Plate Chambers (RPCs) are gaseous detectors 
   capable of delivering a good time resolution below \SI{100}{ps}, 
   at low production cost and displaying magnetic field compatibility.
   In particular, multi-gap RPCs (MRPCs) are able to achieve a time resolution at the level of \SI{50}{ps},
   and therefore they are often used as time-of-flight (TOF) detectors. 
   A use of long pickup strips instead of a pad-based readout enables us to significantly 
   reduce the number of readout electronics and channels. 
   Thus, it leads to avoiding complexity of detector system and reduction of production cost.
   Some groups have developed RPCs with long strips 
   (e.g., with length of \SI{160}{cm} \cite{ref_160}, \SI{180}{cm} \cite{ref_180} and \SI{200}{cm} \cite{ref_200}),
   none of which has been used however in an actual in-beam experiment.

   We developed a \SI{200}{cm}-long strip-MRPC for the LEPS2 experiment.
   LEPS2 aims at studying hadron physics from photo-production reactions 
   at SPring-8, Japan.
   Particles emitted within a polar angle between 5 and 120 degrees are 
   detected by a solenoid spectrometer.
   The MRPC detectors will be installed inside the solenoid magnet bore, 
   which is \SI{2}{m} long and \SI{1.8}{m} in diameter.
   They will be used as TOF detectors for particle identification 
   covering a polar angle region between 30 and 120 degrees.
   The flux of incident particles over the MRPCs is expected to be about \SI{5}{Hz/cm^2}. 
   Therefore, high-rate capability is not required. 
   We will detect charged particles with 99\% efficiency. 
   In the LEPS2 experiment, typical momenta of these particles are below \SI{1}{GeV/c}. 
   The time resolution is required to be better than
    \SI{75}{ps} for $3\sigma$ separation between $\pi/K/p$.
   In previous work \cite{ref_100}, we have shown a time resolution of about \SI{50}{ps} ($\sigma$) 
   at 99\% efficiency for \SI{1}{m}-long strips.
   
   Since a 2m-long strip is about 40 electrical lengths for a typical MRPC signal \cite{ref_diego1}, 
   we pay particular attention to the signal transmission properties of the newly developed detectors.
   Signal transmission along RPC strips has been studied theoretically in the past (e.g. Ref. \cite{ref_trans}).
   In more recent work \cite{ref_diego1}, 
   it was demonstrated that one can obtain a $\times 10$--fold suppression of signal dispersion effects 
   in MRPCs, through the fine balance between inductive and capacitive coupling.
   This situation is called ``electrostatic compensation'' throughout this paper.
   We prepared two types of MRPCs: 
   one with the same layer structure as the previous chamber \cite{ref_100} 
   and the another for which compensation was partially implemented 
   through a subtle modification of the original design.
   We performed a back-to-back comparison of the two types of (test) MRPCs.
   After the encouraging results, we built a (prototype) MRPC relying on the compensation technique, 
   that was characterized exhaustively. 
   Results of both experimental campaigns are reported in this paper.

\section{Mechanical design and readout electronics}
\begin{figure}[tbp]
   \begin{center}
      \includegraphics[clip, width=8.5cm]{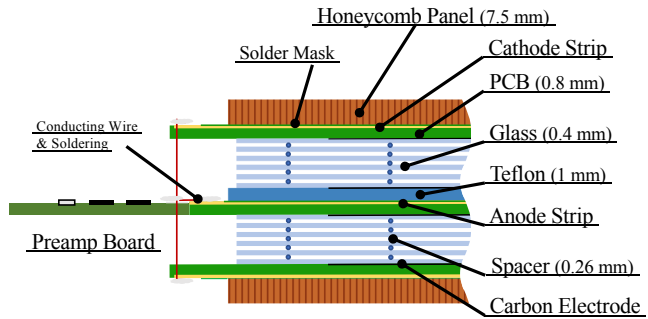}
                 \caption{Schematic of the layer structure of the prototype MRPC for the LEPS2 experiment, 
                   showing the connection between the pre-amplification board and the pickup strip.
                   A Teflon layer has been interleaved in one of the two detector halves, 
                   to bring the detector closer to a compensated state. }
     \label{fig:fig1}
   \end{center}
 \end{figure}

   The layer structure of both the test and prototype MRPCs is 
   very similar to that used in previous detectors \cite{ref_100} 
   (see Fig. \ref{fig:fig1}).
   In the previous work \cite{ref_tomida_jinst}, we optimized the layer structure.
   These MRPCs have 2 stacks and 5 gas gaps per stack.  
   The interval (gas gap) is \SI{0.26}{mm}, and the thickness of glass is \SI{0.4}{mm}.
   The strip dimensions are $200 \times 2.55$ \si{cm^2} and the gap size between adjoining strips is \SI{0.5}{mm}. 
   We changed the end shape of strips from tapered to rectangular.

\begin{figure}[tbp]
   \begin{center}
            \includegraphics[clip, width=5cm]{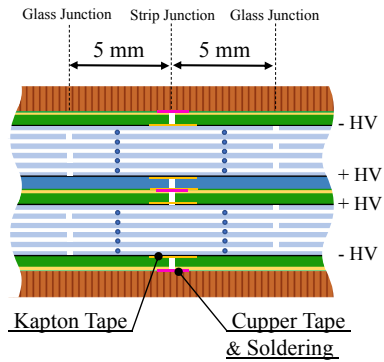}
      \caption{Close-up of the junction region for the 2 pieces of \SI{1}{m} glasses and strip boards, 
                    in the prototype MRPC for the LEPS2 experiment.
                    The positive high voltage is supplied to the two inner carbon electrodes and 
                    the negative one to the outer electrodes. 
                    (The comment about the Teflon layer is identical to that in Fig. \ref{fig:fig1})}
     \label{fig:fig2}
   \end{center}
 \end{figure}

   It is difficult to handle \SI{2}{m}-long thin glasses and PCB boards. 
   Thus, we joined two pieces of \SI{1}{m} glasses and PCB boards. 
   The strips are connected and soldered through short copper tape.
   In order to  prevent discharges at the junction, 
   we assembled the glass layers in such a way that each junction is positioned at $\pm 5$ \si{mm} 
   distance from the center, alternately
   (Fig. \ref{fig:fig2}).

   The MRPCs are operated in avalanche mode. 
   This mode provides a high time resolution, however the induced signal is small. 
   We attached pre-amplification boards directly to both strip ends by soldering short conducting wires, 
   as shown in Fig. \ref{fig:fig1}. 
   Two cascaded chips (RF3776) on the board amplify the signal by about 100 times
    (details are described in Ref. \cite{ref_tomida_jinst2}).
   Each output signal is split in two.
   One output goes to a time-to-digital converter (TDC) through a discriminator, 
   and the other to a charge sensitive analog-to-digital converter (ADC).

\subsection{Test MRPCs}
   In Ref. \cite{ref_diego2}, signal transmission was simulated 
   and the results agree with the experimental results obtained for our previous MRPC structure \cite{ref_100}. 
   Furthermore, in Ref. \cite{ref_diego2}, it was found that 
   interleaving a thin Teflon sheet around the central electrodes 
   would decrease modal dispersion by about a factor $\times 2$.
   In order to study the effect, we built two types of MRPCs:
   one has a G10 layer as the central insulator (identical to \cite{ref_100}) 
   and the other one includes a Teflon layer.

\begin{figure}[tbp]
   \begin{center}
            \includegraphics[clip, width=7cm]{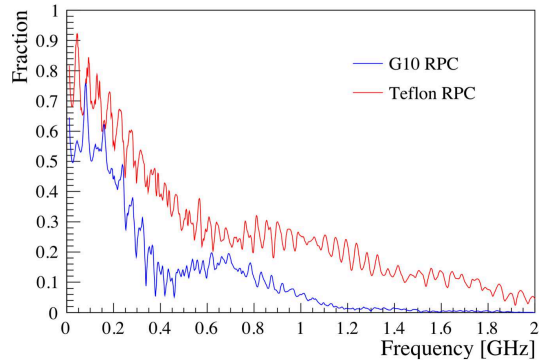}
     \caption{Frequency dependence of the transmitted fraction of signal along one randomly selected MRPC strip, 
                   for the two construction techniques discussed in text. (modulus of the $\text{S}_{21}$ coefficient)
                   }
     \label{fig:fig3}
   \end{center}
 \end{figure}

   We measured the transmitted fraction of signals along the MRPC strips with a Network Analyzer (NWA).
   The results are shown in Fig. \ref{fig:fig3} in the frequency domain.
   One can see the improvement for the Teflon RPC 
   with respect to the earlier design (G10), as expected.
   We also measured the time resolution of these two MRPCs with an electron beam, 
   provided a better signal integrity does not necessarily imply a better time resolution.
   The test results are described in the next section.

\subsection{Prototype MRPCs}
\begin{figure}[tbp]
   \begin{center}
       \includegraphics[clip, width=8.0cm]{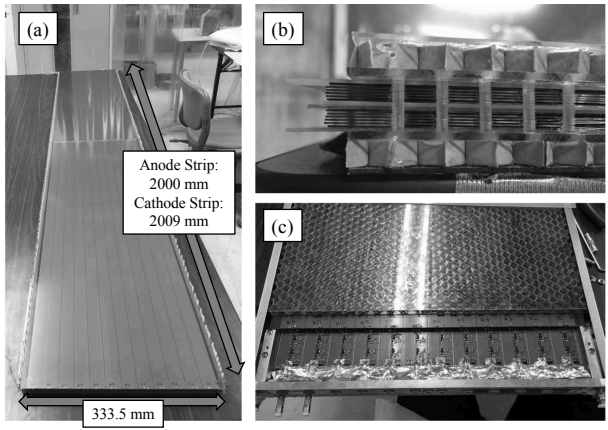}
     \caption{Pictures of the prototype MRPC for the LEPS2 experiment.
                   Figure (a) shows the stack without the top honeycomb layer, viewed from above,
                   and figure (b) shows the strip ends' part (to the left), viewed from the detector side.
                   Figure (c) shows the ends' part of the MRPC and the pre-amplification board.
                   The output part of the board is shielded in order to suppress noise.
                   Both components are in a gas tight chassis.
                   }
     \label{fig:fig_pic}
   \end{center}
 \end{figure}

   The picture of the prototype MRPC is shown in Fig. \ref{fig:fig_pic}.
   The number of strips is 12 and the PCB has \SI{11}{mm} additional free space at both sides,
   therefore its overall width is \SI{333.5}{mm}.
   The length of the anode/cathode PCB is 2000/2009 \si{mm}.
   The dimensions of the glass layer are $333.5 \times 1990$ \si{mm}. 
   Carbon tape is inserted between the PCB and the glass layer as the HV electrode (see Fig. \ref{fig:fig1}). 

   The prototype MRPC follows higher construction standards than the test MRPCs, 
   specially in what concerns detector uniformity, MRPC-FEE connection and noise suppression.
   Based on the results obtained with the test MRPCs,
   we adopted a construction scheme based on a Teflon sheet as the central insulator. 
   In order to increase rigidity and to keep the MRPC uniformity, 
   we added honeycomb layers,  
   which are composed of a honeycomb paper core and two PET panels.
   The MRPC is therefore compressed in a honeycomb sandwich,
   and the distance between the two halves is fixed by screws and nuts.
   Proper grounding to the detector box was essential to keep noise at manageable levels.
   
\section{Test beam measurements with Teflon/G10 MRPCs}
\subsection{Experimental Setup}
   The test was carried out in the SPring-8/LEPS2 beam line. 
   An electron beam was produced by the few-GeV photon beam impinging on a thin Pb converter.
   Two pairs of criss-crossed finger scintillators were placed up and downstream of the MRPC, 
   providing the trigger. 
   The scintillator dimensions were $2 \times 1$ \si{cm^2} upstream 
   and $2 \times 2$ \si{cm^2} downstream.
   The voltage applied to the MRPCs was \SI{13.5}{kV} 
   ($+6.75$ \si{kV} at the inner two carbon electrodes and $-6.75$ \si{kV} at the outer two as shown in Fig. \ref{fig:fig2}).
   The gas mixture consisted of 90\% of Freon (C$_2$H$_2$F$_4$), 5\% of butane and 5\% of SF$_6$. 
   
   To get direct assessment of the signal shape, a DRS4 evaluation board \cite{ref_drs4} was used, 
   before characterizing the detector for time resolution.
   This board can measure signals below \SI{500}{mV} with \SI{500}{MHz} bandwidth.
   With the position of the trigger scintillators fixed, 
   we moved the MRPC along the strip direction to select different hit positions of the electron beam.
   In order to evaluate the signal transmission effect, 
   we studied the dependence of the rise-time 
   \footnote{In this paper, rise-time is defined as the time elapsed from 10\% to 90\% of the signal amplitude.}
   and full width at half maximum (FWHM) of the signals, 
   as a function of the beam position along the detector.

\begin{figure}[tbp]
   \begin{center}
      \includegraphics[clip, width=7.7cm]{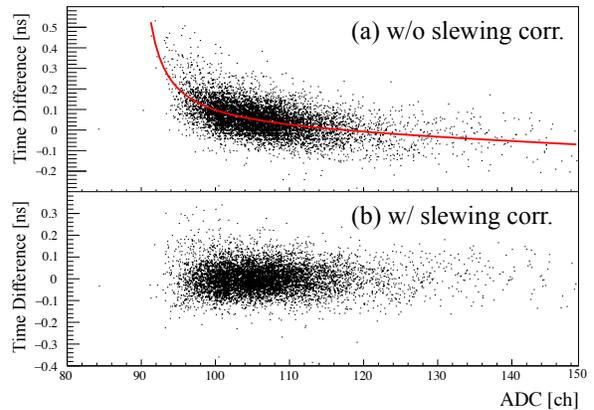}
     \caption{Correlation of the time difference between the MRPC and the RF signal, 
                   and the charge of the signal. The red line in Fig. (a) is a fit.}
     \label{fig:fig4}
   \end{center}
 \end{figure}

   During the standard data taking, the time resolution was evaluated 
   as the r.m.s of a Gaussian fit to the time difference distribution 
   of the MRPC output and the RF signal produced synchronously with the electron bunches
   in the SPring-8 storage ring.
   Since the start time determined from the RF signal has a time jitter below \SI{20}{ps}, 
   the resolution of the time difference is dominated by the one of the MRPC. 
   We used LeCroy 2228A \cite{ref_tdc} and 2249A \cite{ref_adc} 
   to obtain the time and charge information of the induced signals, respectively.
   The jitter of the amplifying, discriminating and digitizing circuit was evaluated to be 15-35 \si{ps}.
   The charge information is used to correct the time-walk effect (slewing correction).
   Figure \ref{fig:fig4} shows the correlation between the time difference and charge with/without slewing correction.
   One can see in Fig. \ref{fig:fig4} (b) that 
   the ADC dependence of the time measurement vanishes after the correction.

  \subsection{Results}
\begin{figure}[tbp]
   \begin{center}
            \includegraphics[clip, width=8.0cm]{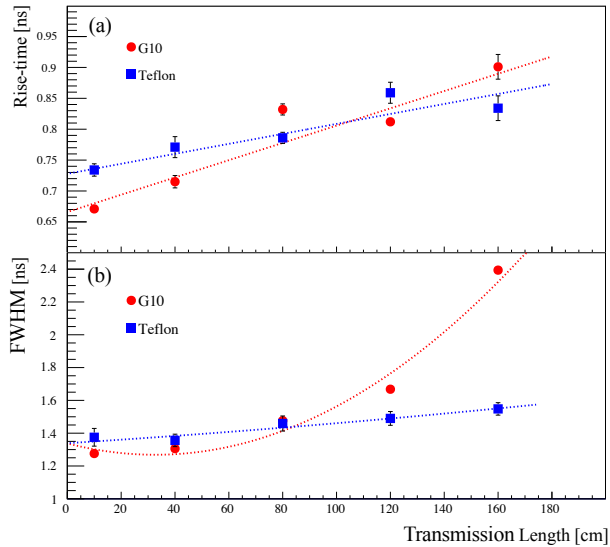}
     \caption{Average rise-time (a) and FWHM (b) of the induced signals 
                   as a function of the beam position along one of the pickup strips of the MRPC.
                   The dotted lines indicate the results of a linear (a) or quadratic (b) fit.
                   }
     \label{fig:fig5}
   \end{center}
 \end{figure}

   The results from the direct waveform inspection 
   are shown in Fig. \ref{fig:fig5}.
   Figure \ref{fig:fig5} (a) shows the position dependence of the signal rise-time and 
   Figure \ref{fig:fig5} (b) shows that of its FWHM, respectively. 
   The horizontal axis represents the distance between the beam position and the readout, 
   which corresponds to the distance over which the induced signal propagates.
   The vertical axis represents the mean value of the rise-time and FWHM of the signals analyzed.
   One can see that, as the distance between the beam and the readout positions increases, 
   both the rise-time and the FWHM increase, more mildly for the Teflon MRPC though.
   These results agree with the fact that the bandwidth of the Teflon MRPC is 
   larger than that of the G10 MRPC, as shown in Fig. \ref{fig:fig3}.
   It supports the fact that modal dispersion is partly suppressed in the Teflon chamber.
   Near the readout position, the G10 MRPC has a significantly smaller rise-time.
   Since the influence of signal transmission is negligible in those conditions, 
   the effect is likely due to channel-to-channel variations of the bandwidth of the pre-amplification board.

    \begin{figure*}[tbp]
   \begin{center}
      \includegraphics[clip, width=14cm]{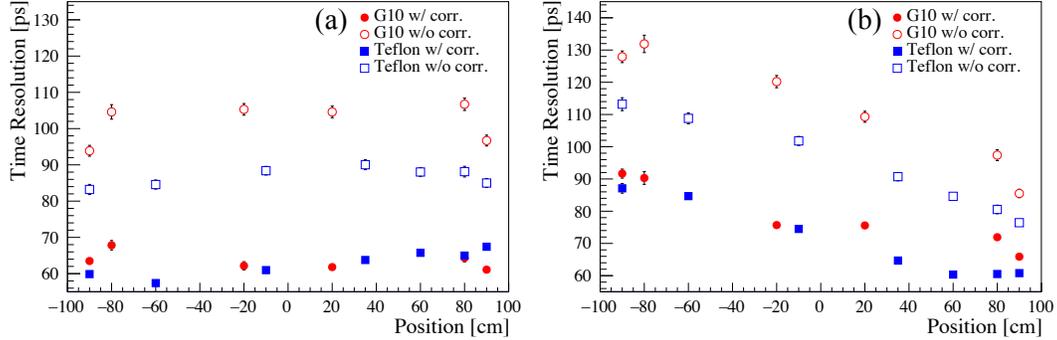}
     \caption{Beam position dependence of the time resolution measured for two types of 
                   test MRPCs developed during the first part of our study.
                   Figure (a) represents the mean time resolution using the signal from both strip ends and 
                   Figure (b) the resolution using only one end. 
                   Closed/open symbols correspond to the resolution with/without slewing correction. 
                   The position \SI{0}{cm} is defined as the center of the strips.}
     \label{fig:fig6}
   \end{center}
 \end{figure*}
 
   Figure \ref{fig:fig6} shows the position dependence of the time resolution, 
   with the center of the strip defined as \SI{0}{cm}.
   Figure \ref{fig:fig6} (a) shows the resolution of 
   the mean time obtained when using signals from both ends of a strip:
   \begin{align}  \label{eq:eq1}
    (t_{\text{left}} + t_{\text{right}} )/2- t_{\text{RF}} 
   \end{align}
   where $t_\text{left(right)}$ is the discriminator firing time 
   of the signal from the $+(-)100$ \si{cm} side and 
   $t_\text{RF}$ is the reference time given by the RF signal from the accelerator.
   Figure \ref{fig:fig6} (b) shows the time resolution obtained from the time information 
   of signals solely from the right end of a strip.
   Closed/open symbols represent the time resolution with/without slewing correction, respectively.
   Either case, the resolution of the one-end readout becomes worse 
   as the distance from the readout point increases, 
   and can be attributed to the transmission effect.
   However, a significant difference between 
   the corrected time resolution of the two type of MRPCs is not seen.
   This result strongly suggests that modal dispersion is the main responsible (but not the only one)
   for the deterioration of the time resolution when signals propagate over long strips.
   Other signal losses (resistive or dielectric) are likely at play \cite{ref_diego1}.
   On the other hand, 
   the Teflon MRPC outperforms a G10-based one in the absence of any type of correction, 
   bringing the time resolution close to \SI{80}{ps}.

\section{Test of LEPS2 prototype MRPC}
   We produced a prototype MRPC, and evaluated its performance.
   Trigger size changed to $2 \times 1$ \si{cm^2} (upstream) and 
   $1.5 \times 1$ \si{cm^2} (downstream).
   The trigger rate was below \SI{50}{Hz}.
   The HV applied to the MRPC was 13 kV which is the optimal value obtained from the HV scan.
   In order to study the effect at the junction and strip ends as well as in the region between strips, 
   we measured the time resolution  through a more thorough scan.
   It is noted that both Teflon detectors in this and previous section are very similar, 
   and differences come down to different quality standards followed during the assembly process.
   
\begin{figure}[tbp]
   \begin{center}
      \includegraphics[clip, width=7.5 cm]{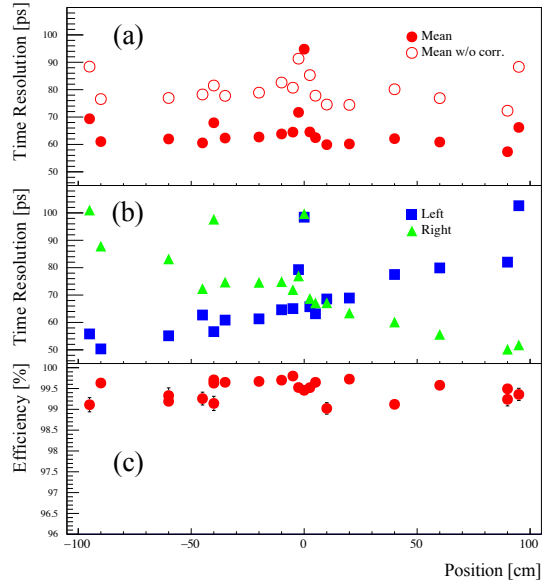}
     \caption{Figure (a) shows the beam position dependence of the mean time resolution
                   for the LEPS2 prototype detector.
                   Closed/open circles correspond to resolution with/without slewing correction. 
                   Figure (b) shows the resolution using one strip end only. 
                   Square/triangle symbols refers, respectively, to the left/right end.
                   Figure (c) shows the detection efficiency.}
     \label{fig:fig7}
   \end{center}
 \end{figure}

   Figure \ref{fig:fig7} (a) shows the position dependence of the mean time resolution and 
   Figure \ref{fig:fig7} (b) shows that when using one strip end only. 
   Closed/open circles indicate the mean time resolution with/without slewing correction, respectively.
   The square/triangle symbols represent 
   the resolution of the one-end readout when the signal is measured at $-100$ \si{cm} ($+100$ \si{cm}).
   As shown in Fig. \ref{fig:fig7} (a),
   using the mean of the time information of both ends of a strip, 
   the position dependence on the resolution is canceled.
   The typical resolutions are \SI{60}{ps} and \SI{80}{ps} with and without slewing correction, respectively.
   Near the edge of the strip, time resolution is a bit worse. 
   This is likely due to the signal reflection (contributing coherently to the far-end signal) and 
   to the relative increase of the differences of the signal transmission path for the near-end signal.
   This deterioration is also seen in previous works \cite{ref_tomida_jinst3}.
   The region with a bad resolution is smaller than that of the previous chamber, 
   however, presumably resulting from the removal of the tapered ends.
   Deterioration of the time resolution is also seen around the junction of the glasses.
   However, this region is limited to a position within $\pm 5$ \si{cm} of the middle of the strip, 
   that will be fiducialized off during physics analysis, similar to the strip end region.
   On the other hand, the resolution of the one-end readout deteriorates smoothly 
   with the beam position.
   These results indicate that, while the junction region shows a deteriorated detector response, 
   it does not affect transmission in a perceptible manner.

   We measured the detection efficiency of the MRPC, too, 
   as a function of the beam position, and is shown in Fig. \ref{fig:fig7} (c).
   It is over 99\% everywhere in the detector.

\begin{figure*}[tbp]
   \begin{center}
      \includegraphics[clip, width=10cm]{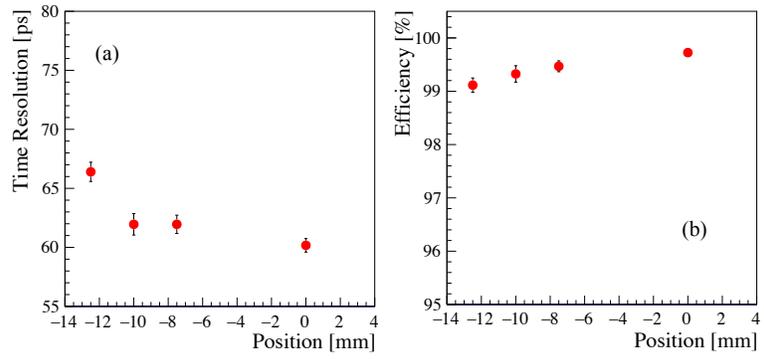}
     \caption{Time resolution (a) and efficiency (b) as a function of the beam position 
                   in the direction perpendicular to the strip.
                   The \SI{0}{cm} is defined as the center of the strip. }
     \label{fig:fig8}
   \end{center}
 \end{figure*}

   Lastly, we also moved the trigger scintillators in a direction perpendicular to the strips.
   These results are shown in Fig. \ref{fig:fig8}.
   In the horizontal axis, \SI{0}{mm} represents the middle of the strip 
   and $- 12.25$ \si{mm} corresponds to the region between adjoined strips. 
   As the beam position moves far from the middle of the strip, 
   the resolution and efficiency become worse.
   However, the resolution is better than \SI{70}{ps} and efficiency is still over 99\% 
   even in the worst case, when the trigger is right in between two strips (half of the beam hits the next strip).

   In addition to the tests described above, we measured the performance of the detector in other cases: 
   $\rm(\hspace{.18em}i\hspace{.18em})$
   with the angle of incidence of the beam with respect to the MRPC changing from \ang{0} to \ang{40}; 
   $\rm(\hspace{.08em}ii\hspace{.08em})$
   with the MRPC placed in the magnetic field; 
   $\rm(i\hspace{-.08em}i\hspace{-.08em}i)$
   with larger trigger area.
   We did not find significant variations of the detector performance in any of those cases.

\section{Summary}
   We developed \SI{2}{m}-long strip-MRPCs for the LEPS2 experiment.
   In order to optimize signal transmission, 
   we changed one of the central layers from G10 to Teflon.
   We found that such a Teflon MRPC comfortably reaches sub-\SI{100}{ps} time resolution 
   before time-charge slewing correction.
   The signal timing characteristics (rise-time and FWHM) are improved accordingly, 
   in agreement with a frequency sweep performed with a NWA.
  
   Based on the Teflon technique we produced a prototype MRPC for LEPS2, 
   respecting higher construction standards, and studying its performance in greater detail.
   The mean time resolution using output signals from 
   both ends of the strip is about \SI{60}{ps} after performing slewing correction.
   The value obtained before any correction, on the other hand, is \SI{80}{ps} 
   that is unprecedented on such long propagation distances, to the best of our knowledge.
   Overall, we found that the achievable time resolution is well below \SI{70}{ps} and 
   the efficiency over 99\%, irrespective of the position (along or across the strip), 
   angle of incidence, magnetic field or trigger area.
   Thus,  we conclude that this MRPC design fulfills the performance requirements of the LEPS2 experiment. 
   We will start mass production of the barrel MRPCs and install them into the solenoid magnet in the near future.

\section*{Acknowledgments}
    The experiments were performed at the BL31LEP of SPring-8 with the approval of the Japan Synchrotron Radiation
    Research Institute (JASRI) as a contract beamline (Proposal No. BL31LEP/6101).
    We thank the SPring-8 staff for providing excellent experimental conditions.
    This work is supported by JSPS KAKENHI (Grant No. 16H06007, 18J22644 and 26287057),
    Osaka University Visiting Scholar Program (Grant No. 16IS004)  
    and the Ministry of Science and Technology of Taiwan.
    D. Gonz\'alez-D\'iaz acknowledges the support from MINECO (Spain) under the  
    Ram\'on y Cajal program (contract RYC-2015-18820).

\section*{References}
{\footnotesize
 \bibliography{mybibfile}
}
\end{document}